\documentclass[fleqn,10pt]{wlscirep}
\title{Optimal array of sand fences} 

\usepackage{graphics}
\usepackage{graphicx}
\usepackage{amssymb}
\usepackage{amsmath}   
\usepackage{color}
\usepackage{xargs}
\usepackage[colorinlistoftodos,prependcaption,textsize=tiny]{todonotes}
\newcommandx{\IZAEL}[2][1=]{\todo[inline,linecolor=red,backgroundcolor=red!25,bordercolor=red,#1]{#2}}
\newcommandx{\ERIC}[2][1=]{\todo[inline,linecolor=blue,backgroundcolor=blue!25,bordercolor=blue,#1]{#2}}
\newcommandx{\EPs}[2][1=]{\todo[linecolor=blue,backgroundcolor=blue!25,bordercolor=blue,#1]{#2}\,}

\author[1]{Izael A. Lima}
\author[1,*]{Asc\^anio D. Ara\'ujo}
\author[2]{Eric J. R. Parteli}
\author[1]{Jos\'e S. Andrade Jr.}
\author[1,3]{Hans J. Herrmann}

\affil[1]{Departamento de F\'isica, Universidade Federal do Cear\'a, 60451-970 Fortaleza, Cear\'a, Brazil} 
\affil[2]{Department of Geosciences, University of Cologne, Pohligstra{\ss}e 3, 50969 Cologne, Germany}
\affil[3]{Computational Physics, IfB, ETH Z\"urich, Schafmattstra{\ss}e~6, 8093 Z\"urich, Switzerland} 
\affil[*]{ascanio@fisica.ufc.br}

\begin{abstract}
Sand fences are widely applied to prevent soil erosion by wind in areas affected by desertification. Sand fences also provide a way to reduce the emission rate of dust particles, which is triggered mainly by the impacts of wind-blown sand grains onto the soil and affects the Earth's climate. Many different types of fence have been designed and their effects on the sediment transport dynamics studied since many years. However, the search for the optimal array of fences has remained largely an empirical task. In order to achieve maximal soil protection using the minimal amount of fence material, a quantitative understanding of the flow profile over the relief encompassing the area to be protected including all employed fences is required. Here we use Computational Fluid Dynamics to calculate the average turbulent airflow through an array of fences as a function of the porosity, spacing and height of the fences. Specifically, we investigate the factors controlling the fraction of soil area over which the basal average wind shear velocity drops below the threshold for sand transport when the fences are applied. We introduce a cost function, given by the amount of material necessary to construct the fences. We find that, for typical sand-moving wind velocities, the optimal fence height (which minimizes this cost function) is around $50\,$cm, while using fences of height around $1.25\,$m leads to maximal cost. 

\end{abstract}

\begin{document}

\flushbottom
\maketitle
\thispagestyle{empty}

\section*{Introduction}

The transport of sand by wind and the concatenated erosion of sediment soils is one of the main causes for the propagation of desertification. Aeolian transport of sand particles is mainly due to saltation, i.e. particles move on approximately ballistic trajectories thereby ejecting new grains upon collision with the soil (splash) \cite{Bagnold_1941}. Moreover, the impacts of sand grains on the soil during saltation are a main factor for the emission of atmospheric dust particles \cite{Shao_et_al_1993} --- which, once entrained, may travel thousands of kilometers in suspension thereby substantially affecting the Earth's climate \cite{Albani_et_al_2014}. To prevent sand transport by wind is thus a concern of broad implication for the society \cite{}.

\begin{figure}[htpb]
\centering
\includegraphics[width=1\linewidth]{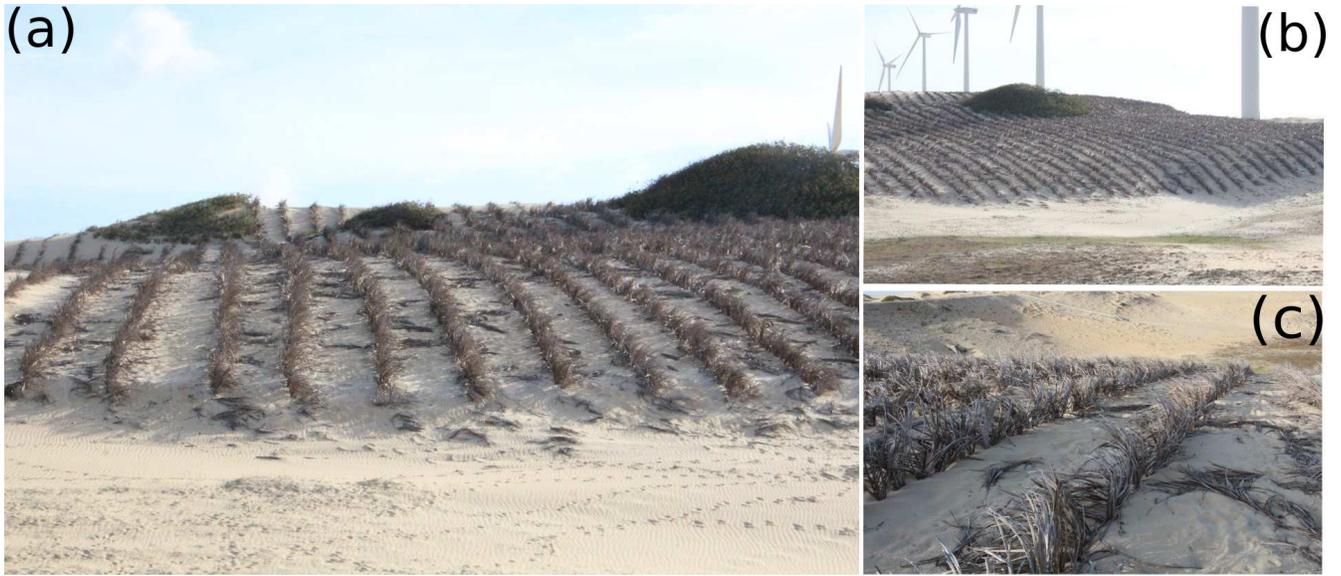}
\caption{{\bf{Application of sand fences to prevent wind erosion --- a field example.}} The images (a)-(c) show fences made of coconut leaves in Paracuru, near Fortaleza, main city of State of Cear\'a in Northeastern Brazil (photo by first authors, I.A.L. and A.D.A.).}
\label{fig:sand_fences}
\end{figure}

Sand fences of various types have been constructed for centuries to control wind erosion and induce dune formation (Fig.~\ref{fig:sand_fences}). Typically, sand fences consist of lightweight wood strips, wire or perforated plastic sheets attached to regularly spaced stakes \cite{Pye_and_Tsoar_1990}. Indeed, the major pre-requisite for a sand fence is that its structure reduces the wind speed, but does not completely block the wind. Indeed, {\em{porous}} fences produce a longer area of leeward sheltered ground than solid fences do --- the latter may also induce strong vortices that extend up to several barrier heights downwind \cite{Cornelis_and_Gabriels_2005}. Notwithstanding the large range of fence designs, all fences operate on the principle to create areas of low wind velocity both in front and behind the fence. In order to protect sand soils from wind erosion, often an array of sand fences is applied, where the fences are erected sequencially at a given spacing along the wind direction \cite{Pye_and_Tsoar_1990}. Depending on the area to be protected a large amount of material may be required to construct the fences. Moreover, the fences must be regularly maintained and replaced due to abrasion of fence's material caused by wind-blown sand. In the present work, we address the problem of predicting the optimal array of fences, that is the array that uses the minimal amount of material necessary to protect a given area of sediment soil from wind erosion. 

Many wind tunnel studies \cite{Baltaxe_1967,Wilson_1987,Lee_and_Kim_1999,Lee_et_al_2002,Xiaoxu_Wu_2013,Guan_2009,Dong_2006,Marijo_Telenta_2014,Ning_Zhang_2015,Tsukahara_et_al_2012}, field works \cite{Savage_1963,Nordstrom_et_al_2012} and numerical simulations \cite{Marijo_Telenta_2014,Hatanaka_et_al_1997,Alhajraf_2004,Wilson_2004,Bouvet_et_al_2006,Santiago_et_al_2007,Benli_Liu_2014,Bitog_et_al_2009} have been performed in order to investigate the characteristics of the turbulent wind flow or sand flux around different types of fences. These studies showed that the amount of sand trapped depends on the fence height, its porosity, the number of fences, their spacing and the wind velocity (for a review see e.g.~Ref.~\cite{Li_and_Sherman_2015}). For sand fences, a porosity of $40\%$ or $50\%$ is recommended since it leads to optimal shielding while avoiding the formation of strong vortices. However, none of these studies focused on adjusting the design of the fence array to reduce building cost. Therefore, we investigate the shear velocity over an array of fences by means of Computational Fluid Dynamic modeling (described below) using the aforementioned porosity values as well as a representative sand-moving wind velocity (defined below). Moreover, we introduce a cost function (presented later in this manuscript), which depends on the fence height and spacing --- to quantify the amount of material needed to construct an array of fences to protect the total area of soil. We will show how this function can be used to obtain the optimal height of an array of sand fences that minimizes the amount of material employed in the fences.

\begin{figure}[htpb]
\centering
\includegraphics[width=1\linewidth]{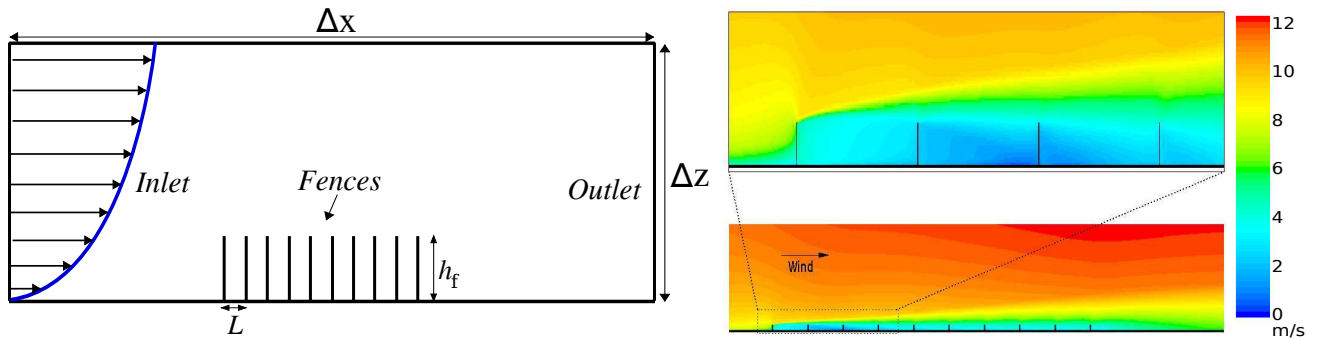}
\caption{{\bf{Numerical experiment.}} {\bf{Left:}} Schematic diagram showing the main quantities characterizing the geometric properties of the wind tunnel with an array of fences. $L$ is the spacing between the fences and $h_{\mathrm{f}}$ their height. The logarithmic wind velocity profile imposed at the inlet is also shown. {\bf{Right:}} Wind velocity magnitude computed using $h_{\mathrm{f}} = 50\,$cm, $L=10\,h_{\mathrm{f}}$ and porosity $\Phi = 50\%$. The wind shear velocity at the inlet is $u_{{\ast}0} = 0.4\,$m$/$s.}
\label{fig:simulation_setup}
\end{figure}

The schematic representation of the setup employed in our calculations is shown in Fig.~\ref{fig:simulation_setup}.
The fences are placed on top of the bottom wall of a two-dimensional channel of height ${\Delta}z = 10\,h_{\mathrm{f}}$ and width ${\Delta}x=80h_{\mathrm{f}}+10L$. Moreover, the soil level in the absence of the fences is considered constant and equal to zero (to simulate an approximately flat sand bed). To avoid that the results are affected by border effects, the dimensions of the box are such that the fences are far enough from the top and side walls of the channel. In other words, we have checked that the results of our calculations do not change significantly if the size of the box is increased. As depicted in Fig.~\ref{fig:simulation_setup}, the wind velocity $u_0(z)$ at the inlet increases with the logarithm of the height $z$ above the bed level ($h$) \cite{Bagnold_1941,Pye_and_Tsoar_1990}, In particular, $u_0(z)=0$ for $z - h = \delta$, where $\delta$ is the surface roughness, and increases with the height above the ground according to the following equation (valid for $z-h \geq \delta$)\cite{Bagnold_1941,Pye_and_Tsoar_1990}: 
\begin{equation}
u_0(z) = \frac{u_{{\ast}0}}{\kappa}{\mathrm{log}}{\frac{z-h}{\delta}}, \label{eq:wind_profile}
\end{equation}
where $\delta$ is the surface roughness, $\kappa = 0.4$ the von K\'arm\'an constant and $u_{{\ast}0}$ the upwind shear velocity of the wind. The shear velocity $u_{{\ast}0}$, which gives the mean (upwind) flow velocity gradient with the height above the soil, is used to define the (upwind) shear stress, 
\begin{equation}
\tau_0 = {\rho}_{\mathrm{air}}{u_{{\ast}0}^2}, \label{eq:tau}
\end{equation}
where ${\rho}_{\mathrm{air}} = 1.225\,$kg$/$m$^3$ stands for air density and $\delta = 100\,{\mu}$m. We note that this value of $\delta$ has been obtained in Ref.~\cite{Almeida_et_al_2006} by fitting Eq.~(\ref{eq:wind_profile}) to the steady-state wind profile within the numerical wind tunnel (Fig.~\ref{fig:simulation_setup}). In Ref.~\cite{Almeida_et_al_2006} this wind profile has been generated by imposing a pressure difference between the in- and outlet of the simulation box, inducing different flow speeds \cite{Almeida_et_al_2006}. Here, we have observed that using other values of surface roughness $\delta$ within the range between $10\,{\mu}$m and $1.0\,$mm \cite{Pye_and_Tsoar_1990} does not change much the shear velocity values obtained in our computations. The boundary conditions, discretization scheme and turbulence model are discussed in detail in Section Methods.

In the CFD simulation, each sand fence is modeled as a vertical, porous wall of height $h_{\mathrm{f}}$, which is varied in the present work from $10\,$cm to $2\,$m. Moreover, each fence consists of a special type of boundary condition which mimics a porous membrane of a certain velocity/pressure drop characteristics \cite{Wilson_1985,Santiago_et_al_2007,Araujo_et_al_2009,Yeh_et_al_2010}. 
Specifically, the pressure drop at height $z$ is given by the equation
\begin{equation}
{\Delta}p(z) = -\frac{1}{4{\Phi}^2}{\rho}_{\mathrm{air}}[u(z)]^2{{\Delta}m}, \label{eq:pressure_drop}
\end{equation}
where $u(z)$ is the wind velocity normal to the fence, that is the horizontal wind speed at height $z$, ${\Delta}m$ is the fence's thickness and $\Phi$ its porosity. In the present work, the fences' thickness is set as ${\Delta}m = 10^{-4}\,$m, while the effect of different values of porosity is investigated.

\section*{Results}

\begin{figure}[htpb]
\centering
\includegraphics[width=1\linewidth]{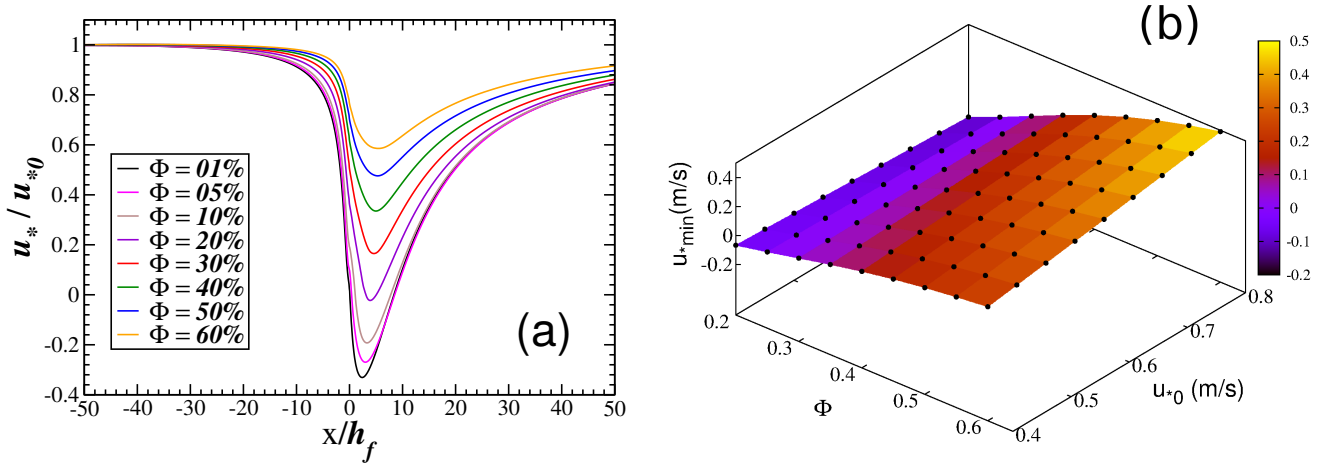}
\caption{{\bf{Results for one single fence. (a)}} Rescaled wind shear velocity, $u_{\ast}/u_{{\ast}0}$, as a function of the rescaled downwind position $x/h_{\mathrm{f}}$. The fence is erected at position $x = 0$. It has height $20\,$cm and different values of porosity according to the legend. Upwind shear velocity is $u_{{\ast}0} = 0.4\,$m$/$s. The minimal value of $u_{\ast}$ in the fence's wake is denoted $u_{{\ast}{\mathrm{min}}}$. {\bf (b)} $u_{{\ast}{\mathrm{min}}}$ as a function of $u_{{\ast}0}$ and $\Phi$ for the same fence height.}
\label{fig:results_one_fence}
\end{figure}

Since most wind-tunnel and field experiments aimed to gain understanding on the effect of fences on sand transport have been performed using one fence, we have started our investigation using one single fence as well. Figure~\ref{fig:results_one_fence} shows results from calculations performed using a fence of height $h_{\mathrm{f}} = 20\,$cm and different values of porosity. The upwind shear velocity in Fig.~\ref{fig:results_one_fence} is $u_{{\ast}0} = 0.4\,$m$/$s, which gives an upwind shear stress of $\tau_0 = 0.196\,$kg$/$m$^2$. As shown previously, this wind shear velocity is a representative value of $u_{\ast}$ above the threshold for saltation in real dune fields \cite{Sauermann_et_al_2003}, although we note that on the field the wind strength has a strongly unsteady behavior and may vary substantially over the time \cite{Ellis_et_al_2012,Sherman_et_al_2013_ICS}.
\vspace{0.5cm}

We see in Fig.\ref{fig:results_one_fence}a the rescaled wind shear velocity $u_{\ast}/u_{{\ast}0}$ as a function of the rescaled downwind position $x/h_{\mathrm{f}}$, for different porosities $\Phi$. The fence is at the position $x = 0$. As we can see, there is a strong decrease of $u_{\ast}$ as the wind approaches the fence from the upwind. This strong decrease is expected as the fence poses an obstacle to the wind thus extracting aeolian momentum. Moreover, the shear velocity decreases further in the fence's wake until a minimum (denoted here $u_{{\ast}{\mathrm{min}}}$) is reached, whereupon upwind flow conditions are recovered later downwind. Similar results have been obtained before experimentally \cite{Zhang_et_al_2010}. We see from Fig.~\ref{fig:results_one_fence}a that very low porosities may lead to strong negative $u_{{\ast}{\mathrm{min}}}$ which means that backward flow occurs in the wake zone. We show in Fig.~\ref{fig:results_one_fence}b the dependence of $u_{{\ast}{\mathrm{min}}}$ on $\Phi$ and $u_{{\ast}0}$. To the best of our knowledge, this is the first time that the three-dimensional diagram of Fig.~\ref{fig:results_one_fence}b is computed. Such a diagram is useful for instance to predict under which conditions of $u_{{\ast}0}$ a fence of given porosity will lead to backward flow or which porosity is necessary to reduce the shear velocity to a pre-determined level below a given $u_{{\ast}0}$.

To get quantitative insight into how to use sand fences for large-scale soil protection, we extend the CFD calculation discussed above to investigate the airflow over an array of many fences. In what follows, we thus focus on the results regarding the array of 10 fences shown in Fig.~\ref{fig:simulation_setup}.

\begin{figure}[t]
\centering
\includegraphics[width=1\linewidth]{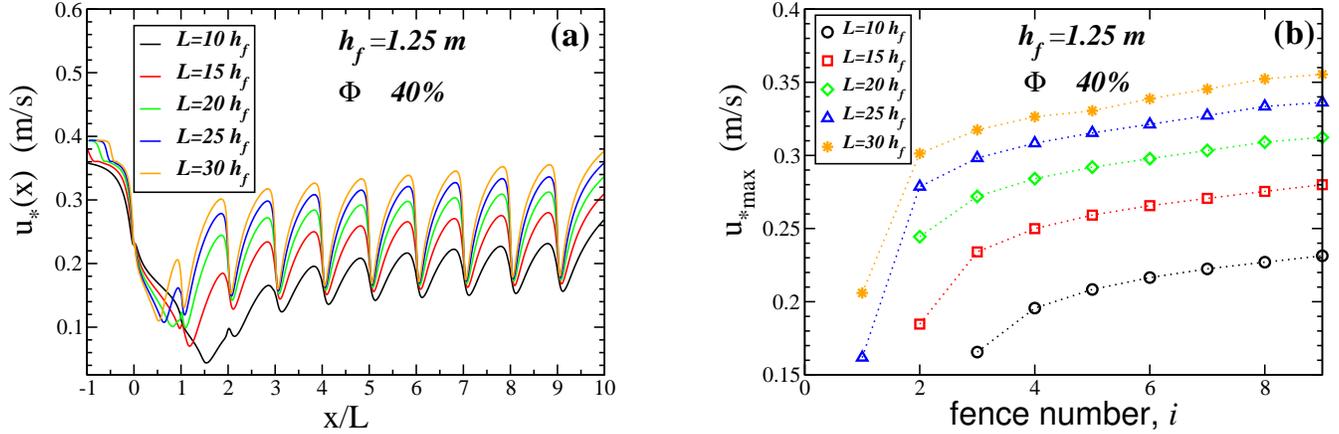}
\caption{{\bf{Wind shear velocity over the array of fences.}} Each fence has height $h_{\mathrm{f}} = 1.25\,$m and porosity $\Phi = 40\%$, while the spacing between the fences, $L$, is varied according to the legend. The wind shear velocity at the inlet is $u_{{\ast}0} = 0.4\,$m$/$s. {\bf(a)} Shear velocity profile with distance downwind and {\bf(b)} maximal values as a function of the fence number, that is between fence $i-1$ and $i$, from second fence ($i = 1$) to the last one ($i = 9$).}
\label{fig:shear_stress}
\end{figure}

Figure \ref{fig:shear_stress} shows the wind shear velocity as a function of the rescaled downwind position $x/L$, where $L$ is the spacing. The first fence is at the position $x = 0$ and there are in total 10 fences at different spacing values (see legend). In the simulation of Fig.~\ref{fig:shear_stress}, the height of the fences is $h_{\mathrm{f}} = 1.25\,$m and the porosity is $\Phi = 40\%$. The wind shear velocity upwind of the fences is $u_{{\ast}0} = 0.4\,$m$/$s.

\begin{figure}[p]
\centering
\includegraphics[width=1\linewidth]{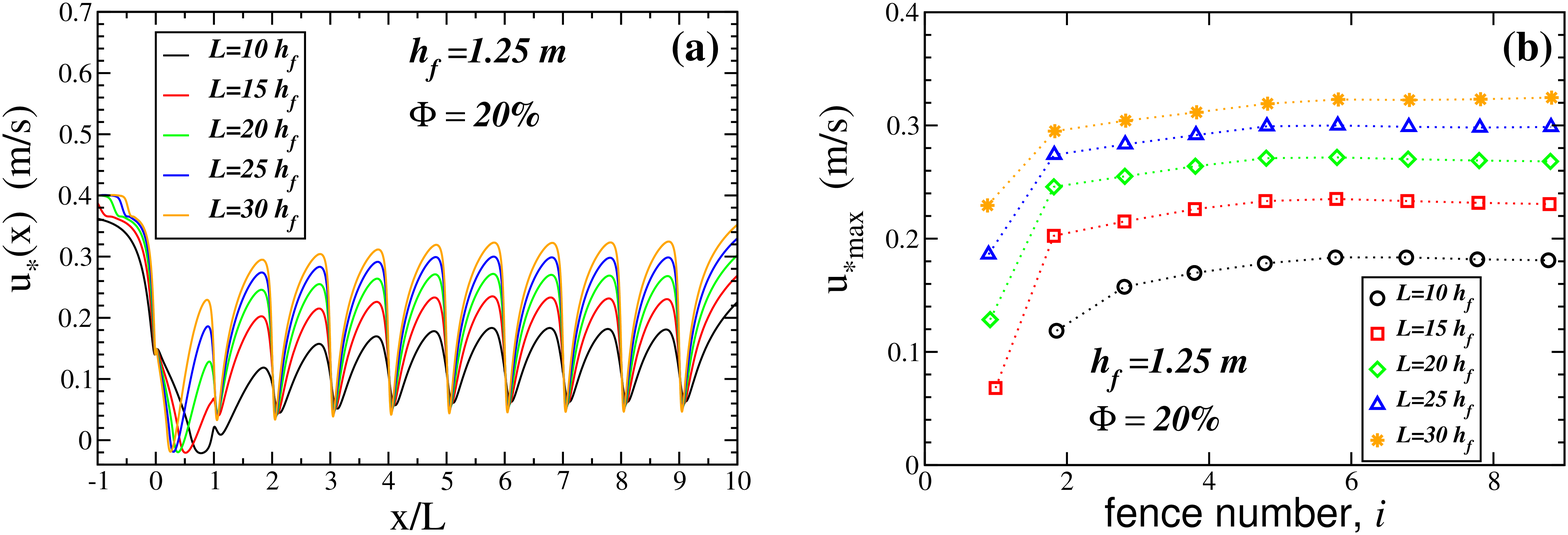}
\caption{{\bf{Wind shear velocity over the array of fences.}} Each fence has height $h_{\mathrm{f}} = 1.25\,$m and porosity $\Phi = 20\%$, while the spacing between the fences, $L$, is varied according to the legend. The wind shear velocity at the inlet is $u_{{\ast}0} = 0.4\,$m$/$s. {\bf(a)} Shear velocity profile with distance downwind and {\bf(b)} maximal values as a function of the fence number, that is between fence $i-1$ and $i$, from second fence ($i = 1$) to the last one ($i = 9$).}
\label{fig:shear_stress_v2}
\end{figure}

As we can see from Fig.~\ref{fig:shear_stress}, after the strong decrease in the shear velocity upwind of the first fence, the behavior of $u_{\ast}(x)$ depends very much on the spacing. In particular, the downwind position at which $u_{{\ast}{\mathrm{min}}}$ is reached after the first fence varies strongly with $L$ (see Fig.~\ref{fig:shear_stress}). Moreover, due to the presence of fences downwind, upwind flow conditions are not achieved within the array. Instead, a maximal wind shear velocity $u_{{\ast}{\mathrm{max}}}$ is reached between each pair of neighbouring fences, whereas $u_{{\ast}{\mathrm{max}}}$ is smaller than the upwind shear velocity $u_{{\ast}0}$. We see that $u_{{\ast}{\mathrm{max}}}$ increases with $L$, which is expected since the larger the fences' spacing the larger the fetch distance available for the wind flow to achieve higher speeds.

Moreover, we see in Fig.~\ref{fig:shear_stress} that $u_{{\ast}{\mathrm{max}}}$ increases substantially from the first to the fourth fence for all values of $L$ investigated. However, further downwind $u_{{\ast}{\mathrm{max}}}$ increases much more slowly distance. We have found, that for a smaller fence porosity ($20\%$), the values of $u_{{\ast}{\mathrm{max}}}$ after the fourth fence are nearly constant with downwind position (that is, with fence number; see Fig.~\ref{fig:shear_stress_v2}). In particular, Figs.~\ref{fig:shear_stress} and \ref{fig:shear_stress_v2} suggest that studies based on one to three fences cannot be used to understand the flow profile over arrays of more fences, as the shear velocity profile over the first three fences is very different from the profile further downwind. We see in Fig.~\ref{fig:shear_stress} that $u_{{\ast}{\mathrm{max}}}$ increases approximately by a factor of two from the second to the last pair of neighbouring fences.

\begin{figure}[htpb]
\centering
\includegraphics[width=0.6\linewidth]{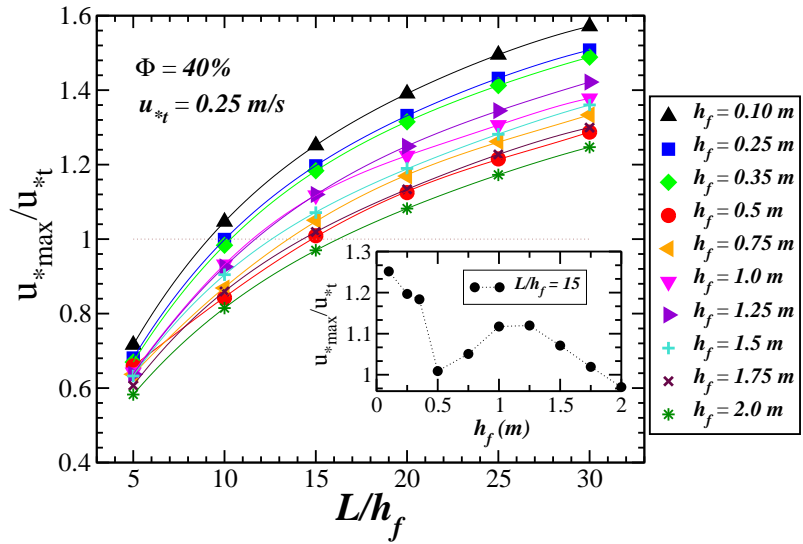}
\caption{{\bf{Maximal wind shear velocity}}, rescaled with the minimal threshold for sustained transport, ${u}_{\ast\mathrm{t}}$, as a function of the rescaled spacing $L/h_{\mathrm{f}}$ for different values of the fences' height $h_{\mathrm{f}}$. Parameters of the simulation are the same as in Fig.~\ref{fig:shear_stress}. The results refer to the value of $u_{\ast\mathrm{max}}$ between the last two fences of the array. The dotted line denotes ${u}_{\ast\mathrm{max}}/{u}_{\ast\mathrm{t}}=1$. The inset shows ${u}_{\ast\mathrm{t}}/u_{{\ast}\mathrm{t}}$ as a function of $h_{\mathrm{f}}$ for fixed value of $L/h_{\mathrm{f}}=15$.}
\label{fig:taumax}
\end{figure}

\begin{figure}[htpb]
\centering
\includegraphics[width=0.6\linewidth]{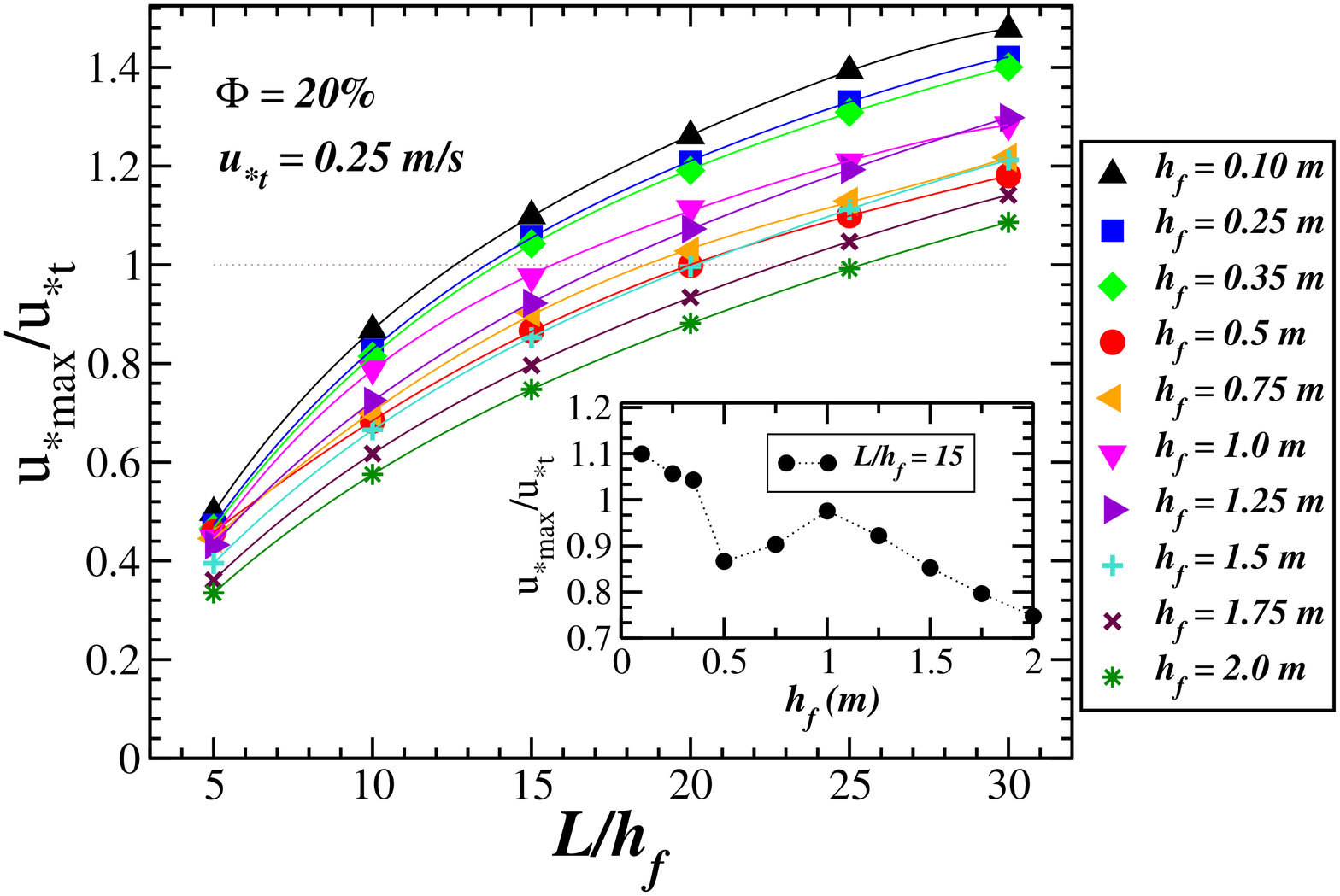}
\caption{{\bf{Maximal wind shear velocity}}, rescaled with the minimal threshold for sustained transport, ${u}_{\ast\mathrm{t}}$, as a function of the rescaled spacing $L/h_{\mathrm{f}}$ for different values of the fences' height $h_{\mathrm{f}}$. Parameters of the simulation are the same as in Fig.~\ref{fig:shear_stress_v2}. The results refer to the value of $u_{\ast\mathrm{max}}$ between the last two fences of the array. The dotted line denotes ${u}_{\ast\mathrm{max}}/{u}_{\ast\mathrm{t}}=1$. The inset shows ${u}_{\ast\mathrm{t}}/u_{{\ast}\mathrm{t}}$ as a function of $h_{\mathrm{f}}$ for fixed value of $L/h_{\mathrm{f}}=15$.}
\label{fig:taumax_v2}
\end{figure}

To investigate the characteristics of the flow over large-scale dune fields in the presence of fences, we now focus on the flow profile far downwind in the fences array. Specifically, we consider the results of the shear velocity between the ninth and tenth fences. Figs.~\ref{fig:taumax} and \ref{fig:taumax_v2} show the value of $u_{{\ast}\mathrm{max}}$, rescaled by the threshold shear velocity $u_{{\ast}\mathrm{t}} = 0.25\,$m$/$s (consistent with medium sand \cite{Pye_and_Tsoar_1990}), as a function of $L/h_{\mathrm{f}}$ for different fences' height, considering a porosity of $40\%$ and $20\%$, respectively. 
We note that $u_{{\ast}{\mathrm{t}}}$ may vary much from field to field depending on soil composition, grain size, the presence of non-erodible elements, humidity and the influence of moisture \cite{Bagnold_1941,Miot_da_Silva_and_Hesp_2010}. From Figs.~\ref{fig:taumax} and \ref{fig:taumax_v2}, it is possible to see for which range of fence height the shear velocity near the surface will not exceed the minimal threshold for transport, that is, ${u}_{{\ast}\mathrm{max}}/u_{{\ast}\mathrm{t}} < 1$, which means total protection of the soil against erosion. The insets in Figs.~\ref{fig:taumax} and \ref{fig:taumax_v2} show ${u}_{\ast\mathrm{max}}/u_{{\ast}\mathrm{t}}$ as a function of $h_{f}$ for a fixed value of $L/h_{\mathrm{f}} = 15$. As we can see, ${u}_{\ast\mathrm{max}}/u_{{\ast}\mathrm{t}}$ displays, in both insets, a maximal value at $h_{f} \approx 1\,$m and a minimum at $h_{f}=0.5\,$m, notwithstanding the different values of porosity associated with each case. Moreover, we note that in the regime of $h_{f} < 0.5\,$m, surface effects become increasingly important as $h_{\mathrm{f}}$ decreases, thus affecting the behavior of ${u}_{\ast\mathrm{max}}/u_{{\ast}\mathrm{t}}$.

\begin{figure}[t]
\centering
\includegraphics[width=1\linewidth]{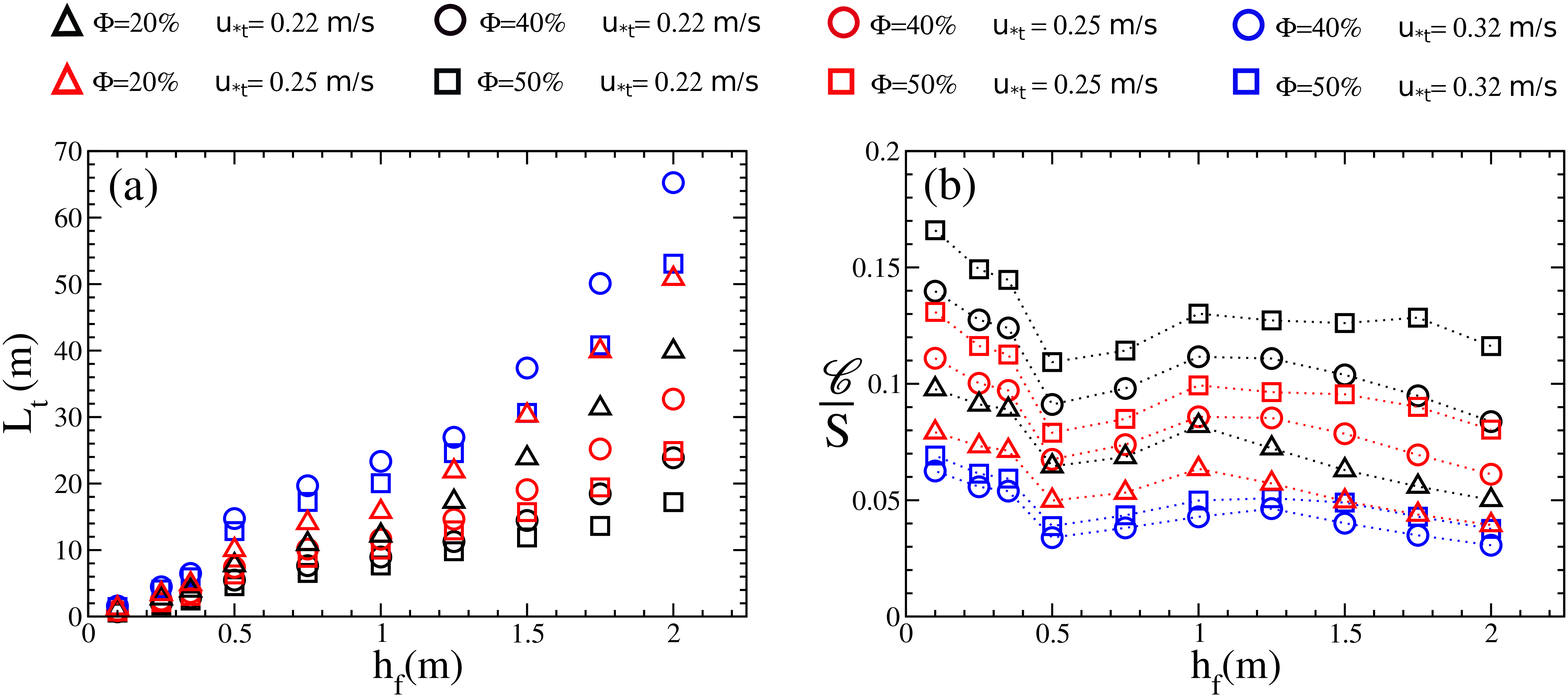}
\caption{{\bf{Obtaining the optimal array of sand fences.}} {\bf(a)} Maximal spacing ensuring no erosion ($u_{{\ast}\mathrm{max}} < u_{{\ast}\mathrm{t}}$) as a function of the fence height $h_{\mathrm{f}}$. {\bf(b)} Cost function divided by target area to be protected (see Eq.~(\ref{eq:cost_function})), as a function of $h_{\mathrm{f}}$. Parameters of the simulation are the same as in Fig.~\ref{fig:taumax}.}
\label{fig:cost_function}
\end{figure}

As we can see from Figs.~\ref{fig:taumax} and \ref{fig:taumax_v2}, the critical value of $L/h_{\mathrm{f}}$ below which total protection against erosion is ensured, that is below which ${u}_{{\ast}\mathrm{max}}/u_{{\ast}\mathrm{t}} < 1$, depends on the fence height. Based on the results of Figs.~\ref{fig:taumax} and \ref{fig:taumax_v2}, we now investigate what is the maximal spacing between fences of given height $h_{\mathrm{f}}$ that guarantees no soil erosion by wind (${u}_{{\ast}\mathrm{max}}/u_{{\ast}\mathrm{t}} < 1$). We call this maximal allowed spacing  $L_{\mathrm{t}}$. The result for $L_{\mathrm{t}}$ is shown in Fig.~\ref{fig:cost_function}a as a function of $h_{\mathrm{f}}$ for several porosities and different wind shear velocity thresholds for sustained transport, $u_{{\ast}{\mathrm{t}}}$. While the values $u_{{\ast}{\mathrm{t}}} = 0.22\,$m$/$s and $0.25\,$m$/$s are consistent with fine and medium sand, respectively, we also performed calculations with a significantly larger $u_{{\ast}{\mathrm{t}}}$ ($0.32\,$m$/$s) to model enhanced resistance to mobilization due to stabilizing agents, such as moisture \cite{Miot_da_Silva_and_Hesp_2010}. We see that $L_{\mathrm{t}}$ increases with $h_{\mathrm{f}}$ regardless of $\Phi$ and $u_{{\ast}{\mathrm{t}}}$, which can be understood by noting that the higher the fences the larger the sheltered distance.

We address the question of which is the most efficient fence array, that is which value of $h_{\mathrm{f}}$ should be used to protect an area of given size under the constraint of using the minimal amount of material to construct the fences. To address this question, we introduce the following {\em{cost function}},
\begin{equation}
\mathcal{C}=\dfrac{S}{L_{t}}h_{\mathrm{f}}, \label{eq:cost_function}
\end{equation}
where $S$ is the total downwind distance of the area to be protected. Note that $S/L_{\mathrm{t}}$ gives the length of the target field in units of numbers of fences. Fig.~\ref{fig:cost_function}b shows the ratio ${\mathcal{C}}/S$. As we can see from this figure, for all studied values of $\Phi$ and $u_{{\ast}{\mathrm{t}}}$, ${\mathcal{C}}/S$ displays a minimum at $h_{\mathrm{f}} \approx 50\,$cm. This fence height is thus the optimal fence height to achieve total protection of a given soil area while ensuring minimal cost. Moreover, we see that, independent of the studied $\Phi$ and $u_{{\ast}{\mathrm{t}}}$,  ${\mathcal{C}}/S$ has a maximum at around $1.25\,$m. This is a surprising result especially considering that the height of fences is often chosen to be $1.0\,$m \cite{Pye_and_Tsoar_1990}. Our result suggests the need for revisiting this choice. Note that our study concerns total protection (no erosion) of a large-scale dune field, where a serial array of multiple fences is used. Our results thus do not apply to an array with less than four fences, because, as shown in Figs.~\ref{fig:shear_stress} and \ref{fig:shear_stress_v2}, the wind shear velocity profile over the first three fences is a transitional one. After the fourth fence, the maximal shear velocity between pairs of neighboring fences is much larger than in the transitional zone and increases only slowly downwind.

We also see in Fig.~\ref{fig:cost_function}b that the value of ${\mathcal{C}}$, for a given $S$, becomes very large as $h_{\mathrm{f}}$ decreases down to values smaller than $50\,$cm. In particular, in this range of small $h_{\mathrm{f}}$, ${\mathcal{C}}$ increases with decreasing $h_{\mathrm{f}}$. This can be explained by the fact that such small fences are rather inefficient for soil protection as their wake region is too short. Moreover, we also see a decrease in ${\mathcal{C}}$ with $h_{\mathrm{f}}$ as the fence height exceeds $1.5\,$m. However, we note that using such large fences is not recommendable as it requires more effort to fixate them in the soil compared to the $50\,$cm ones. We conclude that there is an optimal fence height, which is around $50\,$cm, to guarantee total protection of soils against erosion with a large-scale array of multiple fences.

\section*{Discussion}

We have shown, by means of CFD modeling, that the insights gained from studies using one to three fences cannot be applied to large-scale soil protection where an array of multiple fences is used. Our simulations show that the region corresponding to the first three fences is rather a transitional zone, in which considerable reduction of the shear velocity is achieved. However, after the third fence, the maximal shear velocity between pairs of neighbouring fences is much larger than in the transitional zone and increases only slowly downwind. Future investigations of the effect of sand fences on aeolian erosion for application in large-scale dune fields should thus consider at least four fences. Here, we have considered an array of 10 fences and investigated the shear velocity profile between the last two fences. Our calculations showed that minimal fabrication costs can be achieved if fences with height around 50$\,$cm are used. This is the optimal fence height in an array of multiple fences to guarantee that no erosion occurs in the area to be protected. It is remarkable that this optimal fence height is independent of porosity $\Phi$ and, in particular, of typical threshold wind shear velocity $u_{{\ast}{\mathrm{t}}}$ --- while both parameters affect $L_{\mathrm{t}}$ as shown in Fig.~\ref{fig:cost_function}a. This means that the optimal array of fences applies both for mobile dune sand and for a terrain containing stabilizing elements or moisture.

The present work should be continued by computing the evolution of the sediment landscape in presence of fences. To this end, a morphodynamic modeling tool to simulate aeolian dune formation and migration \cite{Sauermann_et_al_2001,Kroy_et_al_2002,Sauermann_et_al_2003,Duran_et_al_2010,Luna_et_al_2012,Parteli_et_al_2014_EPJST} should be coupled to the CFD simulation in order to model the erosion and deposition patterns resulting from the wind field over the terrain. As a matter of fact, here we have investigated the flow over a flat terrain covered with fences, but local topography evolves in time as sand is deposited in the areas between the fences. Therefore, the results of the present work apply to the initial surface conditions where the terrain is not covered with dunes. Moreover, fences are often applied in combination with the cultivation of vegetation, which acts as sand stabilizer thus helping fixate the soil \cite{Pye_and_Tsoar_1990}. It is thus important to include the effect of plants on the wind flow in future studies as well as to compute the topography resulting from the combined action of wind-blown sand and vegetation by extending the model presented in Refs.~\cite{Duran_and_Herrmann_2006_PRL,Luna_et_al_2009,Luna_et_al_2011} to include the fences.

We remark that, while in our calculations we have assumed constant wind speed, in reality the wind velocity is varying over time which means that occasionally the assumed $u_{\ast} = 0.4\,$m$/$s is exceeded. Calculations using unsteady winds would help to shed light on the characteristics of wind erosion under real conditions. Moreover, our calculations considered the two-dimensional soil profile in longitudinal direction, whereas three-dimensional flow effects \cite{Livingstone_et_al_2007,Melo_et_al_2012} are certainly important if the wind does not hit the fence perpendicularly. Three-dimensional CFD simulations should be thus performed to obtain quantitative insights into such effects. In particular, many different types of fence arrays and other obstacle geometries, such as placing the fences in zig-zag \cite{Bitog_et_al_2009} or checkerboards \cite{Qiu_et_al_2004} are in use. We thus hope that our CFD modeling will inspire future work to calculate the flow over such more complex (three-dimensional) geometries.

\section*{Methods}\label{sec:model}

In the simulations, the fluid (air) is regarded as incompressible and Newtonian, while the average turbulent wind field over the soil is calculated as described in Refs.~\cite{Herrmann_et_al_2005,Araujo_et_al_2013}. The FLUENT Inc. commercial package (version 14.5.7) is adopted to solve the Reynolds-averaged Navier-Stokes equations, whereas in the computations the standard $\kappa-\epsilon$ model is applied to simulate turbulence.

The boundary conditions are modeled as follows. At the inlet of the channel, the logarithmic wind profile (Eq.~(\ref{eq:wind_profile})) is imposed, where the shear velocity $u_{{\ast}0}$ at the inlet is the only parameter of Eq.~(\ref{eq:wind_profile}) varied in the calculations. A constant pressure ($P=0$) is applied at the oulet of the channel in order to produce a pressure gradient in flow direction. Moreover, we apply a non-slip boundary condition to the entire fluid-solid interface comprising the soil and the fences, while the shear stress of the wind at the top wall is set equal to zero (see also Refs.~\cite{Herrmann_et_al_2005,Almeida_et_al_2006,Almeida_et_al_2008,Michelsen_et_al_2015}).

The time-averaged (or Reynolds-averaged) Navier-Stokes equations for the wind flow over the terrain are solved in the fully-developed turbulence regime. The standard $k-\epsilon$ model is used, and the default pressure-velocity coupling scheme (``SIMPLE'') of the solver is applied with its preselected values of parameters, as well as with the default option ``standard wall functions'' (see Ref.~\cite{Araujo_et_al_2013}). In particular, this option applies the wall boundary conditions to all variables of the $k-\epsilon$ model that are consistent with Eq.~(\ref{eq:wind_profile}) along the channel's bottom wall \cite{Launder_and_Spalding_1974}. A second-order upwind discretization scheme is applied to the momentum, whereas for the tubulent kinetic energy and turbulence dissipation rate we apply a first-order upwind scheme \cite{Araujo_et_al_2013}. A square grid with mean spacing of about $0.05h_{\mathrm{f}}$ is applied in the region close to the fence-fluid interface, as well as in the wake region at the front, upward and behind each fence close to the soil. However, this grid is larger in areas which are far away of the fences. 

To solve the transport equations for the standard $k-\epsilon$ model \cite{FLUENT_user_guide}, the following initial conditions are applied: The pressure and velocity are set to zero for all values of $x$ and $z$, while at the left wall ($x=0$), the logarithmic profile Eq.~(\ref{eq:wind_profile}) is imposed. The convergence criteria for the numerical solution of the transport equations are defined in terms of residuals. These residuals provide a measure for the degree up to which the conservation equations are satisfied throughout the flow field. In the present work, convergence is achieved when the normalized residuals for both $k$ and $\epsilon$ fall below $10^{-4}$, and when the normalized residuals for both velocity components fall below $10^{-6}$.

\bibliography{sample}

\begin{thebibliography}{10}
\expandafter\ifx\csname url\endcsname\relax
  \def\url#1{\texttt{#1}}\fi
\expandafter\ifx\csname urlprefix\endcsname\relax\def\urlprefix{URL }\fi
\providecommand{\bibinfo}[2]{#2}
\providecommand{\eprint}[2][]{\url{#2}}

\bibitem{Bagnold_1941}
\bibinfo{author}{Bagnold, R.~A.}
\newblock \emph{\bibinfo{title}{The physics of blown sand and desert dunes}}
  (\bibinfo{publisher}{Methuen, London}, \bibinfo{year}{1941}).

\bibitem{Shao_et_al_1993}
\bibinfo{author}{Shao, Y.}, \bibinfo{author}{Raupach, M.~R.} \&
  \bibinfo{author}{Findlater, P.~A.}
\newblock \bibinfo{title}{Effect of saltation bombardment on the entrainment of
  dust by wind}.
\newblock \emph{\bibinfo{journal}{J. Geophys. Res.}}
  \textbf{\bibinfo{volume}{12}}, \bibinfo{pages}{12,719--12,726}
  (\bibinfo{year}{1993}).

\bibitem{Albani_et_al_2014}
\bibinfo{author}{Albani, S.} \emph{et~al.}
\newblock \bibinfo{title}{{Improved dust representation in the Community
  Atmosphere Model}}.
\newblock \emph{\bibinfo{journal}{Journal of Advances in Modeling Earth
  Systems}} \textbf{\bibinfo{volume}{6}}, \bibinfo{pages}{541--570}
  (\bibinfo{year}{2014}).

\bibitem{Pye_and_Tsoar_1990}
\bibinfo{author}{Pye, K.} \& \bibinfo{author}{Tsoar, H.}
\newblock \emph{\bibinfo{title}{Aeolian sand and sand dunes}}
  (\bibinfo{publisher}{Uwin Hyman, London}, \bibinfo{year}{1990}).

\bibitem{Cornelis_and_Gabriels_2005}
\bibinfo{author}{Cornelis, W.~M.} \& \bibinfo{author}{Gabriels, D.}
\newblock \bibinfo{title}{Optimal windbreak design for wind-erosion control}.
\newblock \emph{\bibinfo{journal}{J. Arid Environ.}}
  \textbf{\bibinfo{volume}{61}}, \bibinfo{pages}{315--332}
  (\bibinfo{year}{2005}).

\bibitem{Baltaxe_1967}
\bibinfo{author}{Baltaxe, R.}
\newblock \bibinfo{title}{{Air Flow Patterns in the Lee of Model Windbreaks}}.
\newblock \emph{\bibinfo{journal}{Arch. Meteorol. Geophys. Bioklimatol. Ser.
  B}} \textbf{\bibinfo{volume}{15}}, \bibinfo{pages}{3} (\bibinfo{year}{1967}).

\bibitem{Wilson_1987}
\bibinfo{author}{Wilson, J.~D.}
\newblock \bibinfo{title}{{On the choice of a windbreak porosity profile}}.
\newblock \emph{\bibinfo{journal}{Boundary-Lay. Meteorol.}}
  \textbf{\bibinfo{volume}{38}}, \bibinfo{pages}{37--49}
  (\bibinfo{year}{1987}).

\bibitem{Lee_and_Kim_1999}
\bibinfo{author}{Lee, S.-J.} \& \bibinfo{author}{Kim, H.-B.}
\newblock \bibinfo{title}{Laboratory measurements of velocity and turbulence
  field behind porous fences}.
\newblock \emph{\bibinfo{journal}{J. Wind Eng. Ind. Aerod.}}
  \textbf{\bibinfo{volume}{80}}, \bibinfo{pages}{311--326}
  (\bibinfo{year}{1999}).

\bibitem{Lee_et_al_2002}
\bibinfo{author}{Lee, S.-J.}, \bibinfo{author}{Park, K.-C.} \&
  \bibinfo{author}{Park, C.-W.}
\newblock \bibinfo{title}{Wind tunnel observations about the shelter effect of
  porous fences on the sand particle movements}.
\newblock \emph{\bibinfo{journal}{Atmos. Environ.}}
  \textbf{\bibinfo{volume}{36}}, \bibinfo{pages}{1453--1463}
  (\bibinfo{year}{2002}).

\bibitem{Xiaoxu_Wu_2013}
\bibinfo{author}{Wu, X.} \emph{et~al.}
\newblock \bibinfo{title}{The effect of wind barriers on airflow in a wind
  tunnel}.
\newblock \emph{\bibinfo{journal}{Journal of Arid Environments}}
  \textbf{\bibinfo{volume}{97}}, \bibinfo{pages}{73--83}
  (\bibinfo{year}{2013}).

\bibitem{Guan_2009}
\bibinfo{author}{Guan, D.-X.} \emph{et~al.}
\newblock \bibinfo{title}{Variation in wind speed and surface shear stress from
  open floor to porous parallel windbreaks: A wind tunnel study}.
\newblock \emph{\bibinfo{journal}{J. Geophys. Res.}}
  \textbf{\bibinfo{volume}{114}}, \bibinfo{pages}{D15106}
  (\bibinfo{year}{2009}).

\bibitem{Dong_2006}
\bibinfo{author}{Dong, Z.}, \bibinfo{author}{Qian, G.}, \bibinfo{author}{Luo,
  W.} \& \bibinfo{author}{Wang, H.}
\newblock \bibinfo{title}{Threshold velocity for wind erosion: the effects of
  porous fences}.
\newblock \emph{\bibinfo{journal}{Environ Geol.}}
  \textbf{\bibinfo{volume}{51}}, \bibinfo{pages}{471--475}
  (\bibinfo{year}{2006}).

\bibitem{Marijo_Telenta_2014}
\bibinfo{author}{Telenta, M.}, \bibinfo{author}{Duhovnik, J.},
  \bibinfo{author}{Kosel, F.} \& \bibinfo{author}{\v{S}ajn, V.}
\newblock \bibinfo{title}{Numerical and experimental study of the flow through
  a geometrically accurate porous wind barrier model}.
\newblock \emph{\bibinfo{journal}{J. Wind Eng. Ind. Aerodyn.}}
  \textbf{\bibinfo{volume}{124}}, \bibinfo{pages}{99--108}
  (\bibinfo{year}{2014}).

\bibitem{Ning_Zhang_2015}
\bibinfo{author}{Zhang, N.}, \bibinfo{author}{Lee, S.~J.} \&
  \bibinfo{author}{Chen, T.-G.}
\newblock \bibinfo{title}{Trajectories of saltating sand particles behind a
  porous fence}.
\newblock \emph{\bibinfo{journal}{Geomorphology}}
  \textbf{\bibinfo{volume}{228}}, \bibinfo{pages}{608--616}
  (\bibinfo{year}{2015}).

\bibitem{Tsukahara_et_al_2012}
\bibinfo{author}{Tsukahara, T.}, \bibinfo{author}{Sakamoto, Y.},
  \bibinfo{author}{Aoshima, D.}, \bibinfo{author}{Yamamoto, M.} \&
  \bibinfo{author}{Kawaguchi, Y.}
\newblock \bibinfo{title}{{Visualization and laser measurements on the flow
  field and sand movement on sand dunes with porous fences}}.
\newblock \emph{\bibinfo{journal}{Exp. Fluids}} \textbf{\bibinfo{volume}{52}},
  \bibinfo{pages}{877--890} (\bibinfo{year}{2012}).

\bibitem{Savage_1963}
\bibinfo{author}{Savage, R.~P.}
\newblock \bibinfo{title}{{Experimental study of dune building with sand
  fences}}.
\newblock In \emph{\bibinfo{booktitle}{Coastal Engineering 1962, Proceedings of
  the 8th International Conference, Mexico City, Mexico.}}
  (\bibinfo{publisher}{American Society of Civil Engineers},
  \bibinfo{year}{1963}).

\bibitem{Nordstrom_et_al_2012}
\bibinfo{author}{Nordstrom, K.~F.}, \bibinfo{author}{Jackson, N.~L.},
  \bibinfo{author}{Freestone, A.~L.}, \bibinfo{author}{Korotky, K.~H.} \&
  \bibinfo{author}{Puleo, J.~A.}
\newblock \bibinfo{title}{Effects of beach raking and sand fences on dune
  dimensions and morphology}.
\newblock \emph{\bibinfo{journal}{Geomorphology}}
  \textbf{\bibinfo{volume}{179}}, \bibinfo{pages}{106--115}
  (\bibinfo{year}{2012}).

\bibitem{Hatanaka_et_al_1997}
\bibinfo{author}{Hatanaka, K.} \& \bibinfo{author}{Hotta, S.}
\newblock \bibinfo{title}{{Finite Element Analysis of Air Flow Around Permeable
  Sand Fences}}.
\newblock \emph{\bibinfo{journal}{Int. J. Numer. Meth. Fl.}}
  \textbf{\bibinfo{volume}{24}}, \bibinfo{pages}{1291--1306}
  (\bibinfo{year}{1997}).

\bibitem{Alhajraf_2004}
\bibinfo{author}{Alhajraf, S.}
\newblock \bibinfo{title}{Computational fluid dynamic modeling of drifting
  particles at porous fences}.
\newblock \emph{\bibinfo{journal}{Environ. Model. Softw.}}
  \textbf{\bibinfo{volume}{19}}, \bibinfo{pages}{163--170}
  (\bibinfo{year}{2004}).

\bibitem{Wilson_2004}
\bibinfo{author}{Wilson, J.~D.}
\newblock \bibinfo{title}{{Oblique, Stratified Winds about a Shelter Fence.
  Part II: Comparison of Measurements with Numerical Models}}.
\newblock \emph{\bibinfo{journal}{J. Appl. Meteorol.}}
  \textbf{\bibinfo{volume}{43}}, \bibinfo{pages}{1392--1409}
  (\bibinfo{year}{2004}).

\bibitem{Bouvet_et_al_2006}
\bibinfo{author}{Bouvet, T.}, \bibinfo{author}{Wilson, J.~D.} \&
  \bibinfo{author}{Tuzet, A.}
\newblock \bibinfo{title}{{Observations and Modeling of Heavy Particle
  Deposition in a Windbreak Flow}}.
\newblock \emph{\bibinfo{journal}{J. Appl. Meteorol. Clim.}}
  \textbf{\bibinfo{volume}{45}}, \bibinfo{pages}{1332--1349}
  (\bibinfo{year}{2006}).

\bibitem{Santiago_et_al_2007}
\bibinfo{author}{Santiago, J.~L.}, \bibinfo{author}{Mart\'in, F.},
  \bibinfo{author}{Cuerva, A.}, \bibinfo{author}{Bezdenejnykh, N.} \&
  \bibinfo{author}{Sanz-Andr\'es, A.}
\newblock \bibinfo{title}{{Experimental and numerical study of wind flow behind
  windbreaks}}.
\newblock \emph{\bibinfo{journal}{Atmos. Environ.}}
  \textbf{\bibinfo{volume}{41}}, \bibinfo{pages}{6406--6420}
  (\bibinfo{year}{2007}).

\bibitem{Benli_Liu_2014}
\bibinfo{author}{Liu, B.}, \bibinfo{author}{Qu, J.}, \bibinfo{author}{Zhang,
  W.}, \bibinfo{author}{Tan, L.} \& \bibinfo{author}{Gao, Y.}
\newblock \bibinfo{title}{Numerical evaluation of the scale problem on the wind
  flow of a windbreak}.
\newblock \emph{\bibinfo{journal}{Scientific Reports}}
  \textbf{\bibinfo{volume}{4}}, \bibinfo{pages}{6619} (\bibinfo{year}{2014}).

\bibitem{Bitog_et_al_2009}
\bibinfo{author}{Bitog, J.~P.} \emph{et~al.}
\newblock \bibinfo{title}{{Numerical simulation of an array of fences in
  Saemangeum reclaimed land}}.
\newblock \emph{\bibinfo{journal}{Atmos. Environ.}}
  \textbf{\bibinfo{volume}{43}}, \bibinfo{pages}{4612--4621}
  (\bibinfo{year}{2009}).

\bibitem{Li_and_Sherman_2015}
\bibinfo{author}{Li, B.} \& \bibinfo{author}{Sherman, D.~J.}
\newblock \bibinfo{title}{{Aerodynamics and morphodynamics of sand fences: A
  review}}.
\newblock \emph{\bibinfo{journal}{Aeolian Research}}
  \textbf{\bibinfo{volume}{17}}, \bibinfo{pages}{33--48}
  (\bibinfo{year}{2015}).

\bibitem{Almeida_et_al_2006}
\bibinfo{author}{Almeida, M.~P.}, \bibinfo{author}{Andrade~Jr., J.~S.} \&
  \bibinfo{author}{Herrmann, H.~J.}
\newblock \bibinfo{title}{Aeolian transport layer}.
\newblock \emph{\bibinfo{journal}{Physical Review Letters}}
  \textbf{\bibinfo{volume}{96}}, \bibinfo{pages}{018001}
  (\bibinfo{year}{2006}).

\bibitem{Wilson_1985}
\bibinfo{author}{Wilson, J.~D.}
\newblock \bibinfo{title}{{Numerical studies of flow through a windbreak}}.
\newblock \emph{\bibinfo{journal}{J. Wind Eng. Ind. Aerod.}}
  \textbf{\bibinfo{volume}{21}}, \bibinfo{pages}{119--154}
  (\bibinfo{year}{1985}).

\bibitem{Araujo_et_al_2009}
\bibinfo{author}{Ara\'ujo, A.~D.}, \bibinfo{author}{Andrade~Jr., J.~S.},
  \bibinfo{author}{Maia, L.~P.} \& \bibinfo{author}{Herrmann, H.~J.}
\newblock \bibinfo{title}{{Numerical Simulation of Particle Flow in a Sand
  Trap}}.
\newblock \emph{\bibinfo{journal}{Granul. Matter}}
  \textbf{\bibinfo{volume}{11}}, \bibinfo{pages}{193--200}
  (\bibinfo{year}{2009}).

\bibitem{Yeh_et_al_2010}
\bibinfo{author}{Yeh, C.-P.}, \bibinfo{author}{Tsai, C.-H.} \&
  \bibinfo{author}{Yang, R.-J.}
\newblock \bibinfo{title}{{An investigation into the sheltering performance of
  porous windbreaks under various wind directions}}.
\newblock \emph{\bibinfo{journal}{J. Wind Eng. Ind. Aerodyn.}}
  \textbf{\bibinfo{volume}{98}}, \bibinfo{pages}{520--532}
  (\bibinfo{year}{2010}).

\bibitem{Sauermann_et_al_2003}
\bibinfo{author}{Sauermann, G.} \emph{et~al.}
\newblock \bibinfo{title}{{Wind velocity and sand transport on a barchan
  dune}}.
\newblock \emph{\bibinfo{journal}{Geomorphology}}
  \textbf{\bibinfo{volume}{54}}, \bibinfo{pages}{245--255}
  (\bibinfo{year}{2003}).

\bibitem{Ellis_et_al_2012}
\bibinfo{author}{Ellis, J.~T.}, \bibinfo{author}{Sherman, D.~J.},
  \bibinfo{author}{Farrell, E.~J.} \& \bibinfo{author}{Li, B.}
\newblock \bibinfo{title}{{Temporal and spatial variability of aeolian sand
  transport: Implications for field measurements}}.
\newblock \emph{\bibinfo{journal}{Aeolian Research}}
  \textbf{\bibinfo{volume}{3}}, \bibinfo{pages}{379--387}
  (\bibinfo{year}{2012}).

\bibitem{Sherman_et_al_2013_ICS}
\bibinfo{author}{Sherman, D.~J.} \emph{et~al.}
\newblock \bibinfo{title}{Characterization of aeolian streamers using
  time-average videography}.
\newblock \emph{\bibinfo{journal}{Journal of Coastal Research}}
  \textbf{\bibinfo{volume}{65}}, \bibinfo{pages}{1331--1336}
  (\bibinfo{year}{2013}).

\bibitem{Zhang_et_al_2010}
\bibinfo{author}{Zhang, N.}, \bibinfo{author}{Kang, J.-H.} \&
  \bibinfo{author}{Lee, S.-J.}
\newblock \bibinfo{title}{{Wind tunnel observation on the effect of a porous
  wind fence on shelter of saltating sand particles}}.
\newblock \emph{\bibinfo{journal}{Geomorphology}}
  \textbf{\bibinfo{volume}{120}}, \bibinfo{pages}{224--232}
  (\bibinfo{year}{2010}).

\bibitem{Miot_da_Silva_and_Hesp_2010}
\bibinfo{author}{Miot~da Silva, G.} \& \bibinfo{author}{Hesp, P.}
\newblock \bibinfo{title}{{Coastline orientation, aeolian sediment transport
  and foredune and dunefield dynamics of Mo\c{c}ambique Beach, Southern
  Brazil}}.
\newblock \emph{\bibinfo{journal}{Geomorphology}}
  \textbf{\bibinfo{volume}{120}}, \bibinfo{pages}{258--278}
  (\bibinfo{year}{2010}).

\bibitem{Sauermann_et_al_2001}
\bibinfo{author}{Sauermann, G.}, \bibinfo{author}{Kroy, K.} \&
  \bibinfo{author}{Herrmann, H.~J.}
\newblock \bibinfo{title}{{A continuum saltation model for sand dunes}}.
\newblock \emph{\bibinfo{journal}{Phys. Rev. E}} \textbf{\bibinfo{volume}{64}},
  \bibinfo{pages}{31305} (\bibinfo{year}{2001}).

\bibitem{Kroy_et_al_2002}
\bibinfo{author}{Kroy, K.}, \bibinfo{author}{Sauermann, G.} \&
  \bibinfo{author}{Herrmann, H.~J.}
\newblock \bibinfo{title}{Minimal model for aeolian sand dunes}.
\newblock \emph{\bibinfo{journal}{Phys. Rev. E}} \textbf{\bibinfo{volume}{66}},
  \bibinfo{pages}{031302} (\bibinfo{year}{2002}).

\bibitem{Duran_et_al_2010}
\bibinfo{author}{Dur\'an, O.}, \bibinfo{author}{Parteli, E. J.~R.} \&
  \bibinfo{author}{Herrmann, H.~J.}
\newblock \bibinfo{title}{A continuous model for sand dunes: Review, new
  developments and application to barchan dunes and barchan dune fields}.
\newblock \emph{\bibinfo{journal}{Earth Surf. Proc. Landforms}}
  \textbf{\bibinfo{volume}{35}}, \bibinfo{pages}{1591--1600}
  (\bibinfo{year}{2010}).

\bibitem{Luna_et_al_2012}
\bibinfo{author}{Luna, M. C. M.~M.}, \bibinfo{author}{Parteli, E. J.~R.} \&
  \bibinfo{author}{Herrmann, H.~J.}
\newblock \bibinfo{title}{Model for a dune field with an exposed water table}.
\newblock \emph{\bibinfo{journal}{Geomorphology}}
  \textbf{\bibinfo{volume}{159-160}}, \bibinfo{pages}{169--177}
  (\bibinfo{year}{2012}).

\bibitem{Parteli_et_al_2014_EPJST}
\bibinfo{author}{Parteli, E. J.~R.}, \bibinfo{author}{Kroy, K.},
  \bibinfo{author}{Tsoar, H.}, \bibinfo{author}{Andrade~Jr., J.~S.} \&
  \bibinfo{author}{P\"oschel, T.}
\newblock \bibinfo{title}{Morphodynamic modeling of aeolian dunes: Review and
  future plans}.
\newblock \emph{\bibinfo{journal}{The European Physical Journal Special
  Topics}} \textbf{\bibinfo{volume}{223}}, \bibinfo{pages}{2269--2283}
  (\bibinfo{year}{2014}).

\bibitem{Duran_and_Herrmann_2006_PRL}
\bibinfo{author}{Dur\'an, O.} \& \bibinfo{author}{Herrmann, H.~J.}
\newblock \bibinfo{title}{Vegetation against dune mobility}.
\newblock \emph{\bibinfo{journal}{Phys. Rev. Lett.}}
  \textbf{\bibinfo{volume}{97}}, \bibinfo{pages}{188001}
  (\bibinfo{year}{2006}).

\bibitem{Luna_et_al_2009}
\bibinfo{author}{Luna, M. C. M.~M.}, \bibinfo{author}{Parteli, E. J.~R.},
  \bibinfo{author}{Dur\'an, O.} \& \bibinfo{author}{Herrmann, H.~J.}
\newblock \bibinfo{title}{Modeling transverse dunes with vegetation}.
\newblock \emph{\bibinfo{journal}{Physica A}} \textbf{\bibinfo{volume}{388}},
  \bibinfo{pages}{4205--4217} (\bibinfo{year}{2009}).

\bibitem{Luna_et_al_2011}
\bibinfo{author}{Luna, M. C. M.~M.}, \bibinfo{author}{Parteli, E. J.~R.},
  \bibinfo{author}{Dur\'an, O.} \& \bibinfo{author}{Herrmann, H.~J.}
\newblock \bibinfo{title}{Model for the genesis of coastal dune fields with
  vegetation}.
\newblock \emph{\bibinfo{journal}{Geomorphology}}
  \textbf{\bibinfo{volume}{129}}, \bibinfo{pages}{215--224}
  (\bibinfo{year}{2011}).

\bibitem{Livingstone_et_al_2007}
\bibinfo{author}{Livingstone, I.}, \bibinfo{author}{Wiggs, G. F.~S.} \&
  \bibinfo{author}{Weaver, C.~M.}
\newblock \bibinfo{title}{{Geomorphology of desert sand dunes: A review of
  recent progress}}.
\newblock \emph{\bibinfo{journal}{Earth-Sci. Rev.}}
  \textbf{\bibinfo{volume}{80}}, \bibinfo{pages}{239--257}
  (\bibinfo{year}{2007}).

\bibitem{Melo_et_al_2012}
\bibinfo{author}{Melo, H. P.~M.}, \bibinfo{author}{Parteli, E. J.~R.},
  \bibinfo{author}{Andrade, J.~S.} \& \bibinfo{author}{Herrmann, H.~J.}
\newblock \bibinfo{title}{Linear stability analysis of tranverse dunes}.
\newblock \emph{\bibinfo{journal}{Physica A}} \textbf{\bibinfo{volume}{391}},
  \bibinfo{pages}{4606--4614} (\bibinfo{year}{2012}).

\bibitem{Qiu_et_al_2004}
\bibinfo{author}{Qiu, G.~Y.}, \bibinfo{author}{Lee, I.-B.},
  \bibinfo{author}{Shimizu, H.}, \bibinfo{author}{Gao, Y.} \&
  \bibinfo{author}{Ding, G.}
\newblock \bibinfo{title}{Principles of sand dune fixation with straw
  checkerboard technology and its effects on the environment}.
\newblock \emph{\bibinfo{journal}{J. Arid Environ.}}
  \textbf{\bibinfo{volume}{56}}, \bibinfo{pages}{449--464}
  (\bibinfo{year}{2004}).

\bibitem{Herrmann_et_al_2005}
\bibinfo{author}{Herrmann, H.~J.}, \bibinfo{author}{Andrade~Jr., J.~S.},
  \bibinfo{author}{Schatz, V.}, \bibinfo{author}{Sauermann, G.} \&
  \bibinfo{author}{Parteli, E. J.~R.}
\newblock \bibinfo{title}{Calculation of the separation streamlines of barchans
  and transverse dunes}.
\newblock \emph{\bibinfo{journal}{Physica A}} \textbf{\bibinfo{volume}{357}},
  \bibinfo{pages}{44--49} (\bibinfo{year}{2005}).

\bibitem{Araujo_et_al_2013}
\bibinfo{author}{Ara\'ujo, A.~D.}, \bibinfo{author}{Parteli, E. J.~R.},
  \bibinfo{author}{P\"oschel, T.}, \bibinfo{author}{Andrade~Jr., J.~S.} \&
  \bibinfo{author}{Herrmann, H.~J.}
\newblock \bibinfo{title}{Numerical modeling of the wind flow over a transverse
  dune}.
\newblock \emph{\bibinfo{journal}{Scientific Reports}}
  \textbf{\bibinfo{volume}{3}}, \bibinfo{pages}{2858} (\bibinfo{year}{2013}).

\bibitem{Almeida_et_al_2008}
\bibinfo{author}{Almeida, M.~P.}, \bibinfo{author}{Parteli, E. J.~R.},
  \bibinfo{author}{Andrade~Jr., J.~S.} \& \bibinfo{author}{Herrmann, H.~J.}
\newblock \bibinfo{title}{Giant saltation on {M}ars}.
\newblock \emph{\bibinfo{journal}{Proceedings of the National Academy of
  Sciences}} \textbf{\bibinfo{volume}{105}}, \bibinfo{pages}{6222--6226}
  (\bibinfo{year}{2008}).

\bibitem{Michelsen_et_al_2015}
\bibinfo{author}{Michelsen, B.}, \bibinfo{author}{Strobl, S.},
  \bibinfo{author}{Parteli, E. J.~R.} \& \bibinfo{author}{P\"oschel, T.}
\newblock \bibinfo{title}{Two-dimensional airflow modeling underpredicts the
  wind velocity over dunes}.
\newblock \emph{\bibinfo{journal}{Scientific Reports}}
  \textbf{\bibinfo{volume}{5}}, \bibinfo{pages}{16572} (\bibinfo{year}{2015}).

\bibitem{Launder_and_Spalding_1974}
\bibinfo{author}{Launder, B.~E.} \& \bibinfo{author}{Spalding, D.~B.}
\newblock \bibinfo{title}{The numerical computation of turbulent flows}.
\newblock \emph{\bibinfo{journal}{Comput. Method. Appl. M.}}
  \textbf{\bibinfo{volume}{3}}, \bibinfo{pages}{269--289}
  (\bibinfo{year}{1974}).

\bibitem{FLUENT_user_guide}
\bibinfo{author}{{FLUENT Inc.}}
\newblock \emph{\bibinfo{title}{Standard, RNG, and Realizable Models Theory}}
  (\bibinfo{year}{2006}).
\newblock \bibinfo{note}{FLUENT 6.3 User's Guide, Section 12.4.
  \url{http://aerojet.engr.ucdavis.edu/fluenthelp/html/ug/node477.htm}}.

\end{thebibliography}

\section*{Acknowledgements}

This work was supported in part by CAPES, CNPq and FUNCAP (Brazilian agencies), the Brazilian Institute INCT-SC, by the German Research Foundation (DFG) Grant RI 2497/3-1 and by ERC Advanced grant FP7-319968 FlowCCS of the European Research Council.

\section*{Author contributions statement}

I.A.L., A.D.A., E.J.R.P., J.S.A.Jr and H.J.H. designed the research, analyzed the data and wrote the paper.

\section*{Additional information}

Competing financial interests: The authors declare no competing financial interests.


\end{document}